\documentclass[12pt,a4paper]{article}
\usepackage{amsmath,amsthm,amsfonts}
\newcommand{\der}[1]{\frac{d}{d #1 }}

\author{Israel Klich}
\begin{document}
\title{Casimir's energy of a conducting sphere
and of a dilute dielectric ball} \maketitle 

\begin{center}
Departments of Applied Mathematics and Physics, Technion, 32000
Haifa, Israel\footnote{klich@tx.technion.ac.il}
\end{center}
\begin{abstract}
In this paper we sum over the spherical modes appearing in the
expression for the Casimir energy of a conducting sphere and of a
dielectric ball (assuming the same speed of light inside and
outside), before doing the frequency integration. We derive closed
integral expressions that allow the calculations to be done to all
orders, without the use of regularization procedures. The
technique of mode summation using a contour integral is critically
examined.
\end{abstract}
\bigskip
\section{Introduction}
The Casimir effect is a remarkable consequence of QFT, exhibiting
the reality of the zero point energy of the vacuum. This energy
can be made manifest by studying its dependence on various
constraints imposed on the electromagnetic field. Casimir's
\cite{casimir48} original analysis for parallel plates yielded a
force per unit area  ($\hbar=c=1$)
\begin{equation}
-\frac{\pi^2}{240 a^4}
\end{equation}
between two conducting plates separated by a distance $a$. The
Casimir force for a conducting spherical shell was first
calculated by Boyer \cite{boyer68}. The result was indeed
puzzling: the force turned out to be repulsive. This was
surprising in view of the connection between Casimir's force and
Van-Der-Waals forces which are attractive. This phenomenon may be
also related to a fundamental difference between the spherical
Casimir problem and that of the plates: expansion of the sphere
changes the distances between the points on the shell itself, and
thus it is not clear weather the Casimir energy (associated with
interaction among different patches on the shell) and
electromagnetic the self energy of the shell can be separated
\cite{nussinov}. It is not surprising, however, that this in turn
called for a large number of subsequent calculations by various
techniques
\cite{davies72,balian78,milton78,nesterenko97,barton99}. There has
been some controversy concerning the regularization procedures
used in the calculations, bearing, for example on the relevance of
Casimir's effect to sonoluminescence
\cite{brevik99,liberati1,liberati2,milton98}. Since the repulsive
nature of the Casimir force for spherical boundary conditions is
somewhat counterintuitive it seems of importance to justify it
more rigorously.

    We define the Casimir energy of a conducting
sphere as follows: Let $\omega_{n}^{(a,R)}$ be the eigenmodes of
the electromagnetic field inside two concentric spherical shells
of radii $a$ and $R$
 $(a<<R)$, where a large sphere of radius $R$ is introduced as an
infrared cutoff in order to discretize the spectrum outside the
smaller sphere as well as inside it. The zero point energy of this
system, namely $\frac{1}{2}\sum_n \omega_{n}^{(a,R)}$ is a
divergent sum. The Casimir energy is then defined as the zero
point energy difference between our configuration and a
configuration where the inner sphere is of radius $b$ ($a<<b<R$)
 \cite{plunien86}\footnote{this form of subtraction has the advantage of having
a natural one to one correspondence between the eigenmodes of the
constrained system (inner radius $a$) to those of the
``unconstrained" one (where $b$ and $R$ are taken to $\infty$) }

\begin{equation}
E_{C}=\lim_{\sigma\rightarrow
0}\lim_{b\rightarrow\infty}\lim_{R\rightarrow\infty}
\frac{1}{2}\sum_{n}\big( e^{-\sigma \omega_{n}^{(a,R)}}
\omega_{n}^{(a,R)}-e^{-\sigma \omega_{n}^{(b,R)}}
\omega_{n}^{(b,R)}\big)
\end{equation}
Here $e^{-\sigma \omega_n}$ is an exponential regulator. We show
that if the subtraction procedure is done properly, no further
regularization (apart from keeping the naive exponential
regulator, required to make the sums well defined, and the
subtraction) is actually needed. In the sequel , we follow closely
the procedure of mode summation using a contour integral, but our
main result can also be applied to the Green's function method
 \cite{milton78}.

    Most of the calculations have been carried out
for the case of a conducting sphere and of a dielectric ball
having the same speed of light inside and outside, that is,
$\sqrt{\epsilon\mu}=\sqrt{\epsilon'\mu'}=1$ \cite{brevik82}, where
$\epsilon,\mu$ are the permittivity and permeability of the medium
surrounding a ball of radius $a$, and $\epsilon',\mu'$ are the
permittivity and permeability inside the ball. It has been shown
that the Casimir energy in the latter case is given by
\cite{nesterenko97,brevik98}
\begin{equation}
E=-\frac{1}{2\pi a}\int_{-\infty}^{\infty} dy
 e^{iy\delta} \sum_{l=1}^{\infty}(l+\frac{1}{2}) x\der{x}
\ln{(1-\xi^{2} {(s_{l}e_{l})'}^{2})}
\end{equation}
where $x=|y|, \nu=l+\frac{1}{2}$ and
\begin{equation}
\xi=\frac{\mu-\mu'}{\mu+\mu'}; \quad \delta=\frac{\tau}{a}
\end{equation}
\begin{equation}
s_{l}(x)=ixj_l(ix)=\sqrt{\frac{\pi
x}{2}}e^{-\frac{i\pi\nu}{2}}J_{\nu}(ix)
\end{equation}
\begin{equation}
e_{l}(x)=ixh_l^{(1)}(ix)=i\sqrt{\frac{\pi
x}{2}}e^{\frac{i\pi\nu}{2}} H^1_{\nu}(ix)
\end{equation}
and in the end of the calculation the regulator $\delta$ is taken
to zero. In the case $\xi=1$, (3) is equivalent to the expression
for the energy of a conducting sphere in vacuum.

    In order to evaluate (3), extensive use has been made
of uniform asymptotic approximations of the modified Bessel
functions $s_{l}(x)$ and $e_{l}(x)$. Our evaluation refrains from
the use of these asymptotics, which, although very convenient in
actual calculations, somewhat obscures the nature of the
expression (3). By studying the $l$-sum of this expression before
making the frequency integration, using simple orthogonality
considerations we arrive at a general formula for sums of the form
$\sum(2l+1)(s_{l}e_{l})^{n}$. We show that to first order in
$\xi^2$ the integrand consists of a constant term (which is
independent of the radius) and a function of $a$. Thus the
divergence appearing in the evaluation of (3) is due to
integration of this constant over all frequencies. This divergence
was already pointed out by Candelas \cite{candelas82}, however,
this constant term would not be present if subtraction of the
Casimir energy of a large radius sphere is done properly.

    We start by carefully reviewing the procedure of mode summation,
using a contour integral, as was done in some previous works
\cite{bordag96,nesterenko97,brevik98}. In this method the final
integration is carried along the imaginary axis. Justification for
the neglect of the part of the contour away from this axis will be
given, by concrete estimates, taking into account the existence of
an infinite number of poles of the integrand on the real axis.

\section{Mode summation for conducting spherical shells}
In order to make the discussion well defined we investigate the
following situation \cite{bowers99}: A conducting sphere of radius
$a$ is placed (concentrically) inside a large conducting sphere of
radius $R$. Using multipole expansion \cite{jackson75} the
appropriate modes of the electromagnetic fields can be found,
through the boundary conditions $B_r=0$ and $E_{\theta,\varphi}=0$
on the shells. The eigenmodes are given as solutions to the
equations:
\\ (I) Inside the inner sphere
\begin {equation}
\Delta_l^{(1)}\equiv j_l(\omega a)=0 
\end
{equation}
\begin {equation}
\Delta_l^{(2)}\equiv {d\over dr}[rj_l(\omega r)]\biggr|_{r=a}=0
\end {equation}
II) Between the shells
\begin {equation}
\Delta_l^{(3)}\equiv j_l(\omega R)n_l(\omega a)-j_l(\omega
a)n_l(\omega R)=0  
\end {equation}
\begin {equation}
\Delta_l^{(4)}\equiv {d\over dr}[rj_l(\omega
r)]\biggr|_{r=a}{d\over dr}[rn_l(\omega r)]\biggr|_{r=R}-{d\over
dr}[rj_l(\omega r)]\biggr|_{r=R}{d\over dr}[rn_l(\omega
r)]\biggr|_{r=a}=0
\end{equation}
Where equations (7),(9) refer to transverse electric modes, and
(8),(10) refer to transverse magnetic modes. Next we define the
function \cite{bowers99}
\begin{equation}
\Delta_{l}(\omega;a,R)=\omega^{2}\Delta_l^{(1)}\Delta_l^{(2)}\Delta_l^{(3)}\Delta_l^{(4)}
\end{equation}
$\Delta_{l}$ is an analytic function, whose positive real zeros
are the modes of the electromagnetic field inside the concentric
spheres.

    In order to sum over the modes of a given $l$, we use the identity
\cite{nesterenko97,bordag96}
\begin{equation}
\sum_{n=1}^{N} \omega_{(n,l)} e^{-\sigma \omega_{(n,l)}}={1\over
2\pi i}\oint_{C_{N}}dz e^{-\sigma z}z{d\over dz}\ln \Delta_{l}(z)
\end{equation}
Here $C_{N}$ is a contour which encircles the zeros
$\omega_{(1,l)}, .. \omega_{(N,l)}$ of $\Delta_{l}$ . Since there
is an infinite number of modes for each $l$, this sum will diverge
if we take $N$ to infinity and $\sigma$ to zero. As usual, the
Casimir energy should be identified with that obtained after the
subtraction of the zero point energy of the unconstrained ``free"
system \cite{plunien86},
\begin{equation}
\Delta E(\sigma,b,R)=\lim_{N\rightarrow\infty}{1\over 2\pi
i}\sum_{l=1}^{\infty} (l+{1\over 2})\oint_{C_{N}} dz e^{-\sigma
z}z F_{l} \label{deltae}
\end{equation}
Where
\begin{equation}
F_{l}(z;a,b,R)=\Big({d\over dz}\ln \Delta_{l}(z;a,R)-{d\over
dz}\ln \Delta_{l}(z;b,R)\Big).
\end{equation}
The Casimir energy is then given by the limit:
\begin{equation}
E_C=\lim_{\sigma\rightarrow
0}\lim_{b\rightarrow\infty}\lim_{R\rightarrow\infty} \Delta
E(\sigma,b,R).
\end{equation}
\begin{figure}\center {\input epsf \epsfbox{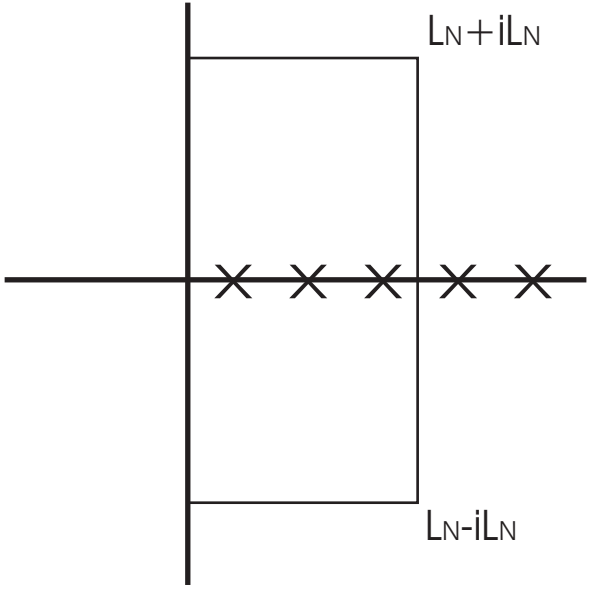}}
\caption{A rectangular contour in the complex plane.} \label{f1}
\end{figure}
    After the limits $R\rightarrow\infty$ and $b\rightarrow\infty$ are
taken the remaining expression will have no reference to $R$ and
$b$. Removal of the ultra-violet cut off by taking the limit
$\sigma\rightarrow0$ is done in the end of the derivation, after
insuring convergence for all values of $b$ and $R$.

    In order to make the evaluation of the contour integrals easier we
take rectangular contours of side $L_{N}$ (Fig. 1). Our first aim
is to carry the entire integration along the imaginary axis, by
showing that the contribution of the part of the contour away from
it vanishes as $L_N$ goes to infinity. The justification of this
neglect, however, is not obvious, since the integrand in
(\ref{deltae}) has infinitely many singularities along the real
axis, which become ever denser as $R\rightarrow\infty$. This
follows from the property that for any finite $R$ there is an
infinite number of field modes, and subsequently there is a
corresponding infinite number of poles of $z{d\over dz}\ln
\Delta_{l}(z,b,R)$ along the real axis. Even after taking the
limit $R\rightarrow\infty$ the spectrum will still consist of a
discrete set (eigenmodes inside the inner shell) imbedded in a
continuum part. Indeed, let us examine the behavior of
$\Delta_{l}(\omega;a,R)$ for large arguments $\omega$. Using the
asymptotic behavior \cite{jackson75}
\begin{eqnarray}
j_{l}\rightarrow\frac{1}{x}\sin(x-\frac{l\pi}{2})\\
n_{l}\rightarrow-\frac{1}{x}\cos(x-\frac{l\pi}{2})
\end{eqnarray}
we obtain
\begin{equation}
\Delta_{l}(\omega;a,R)\rightarrow\frac{1}{2a^{2}R\omega}
\sin(2\omega a-l\pi)\sin^{2}(\omega(R-a)) \label{asym}
\end{equation}
Thus for large arguments the modes behave as $\frac{n\pi}{2a}$ or
$\frac{n\pi}{2(R-a)}$ where $n$ is an integer. (Note that the
density of modes for large frequencies is similar to that
appearing in the one dimensional version of the standard Casimir
effect for parallel conducting plates separated by a distance $a$
inside a large box of length $R$.) Let us examine the behavior of
the integrand for large values of $|\omega|$. Substituting the
asymptotics (\ref{asym}), we show that the integral on a contour
that does not pass through one of the poles, decays exponentially
fast. The integrand is then simply:
\begin{eqnarray}
&&F_{l}(\omega;a,b,R)=2\big(a(\cot(2\omega
a-l\pi)-\cot(\omega(R-a)))+\nonumber\\&&
R(\cot(\omega(R-a))-\cot(\omega(R-b)))+\nonumber\\&&
b(\cot(\omega(R-a))-\cot(2\omega b-l\pi))\big)
\end{eqnarray}
It is easy to check that for ${\rm Im}~\omega\rightarrow\infty$
and any real $A,B$,
\begin{equation}
|\cot(A \omega)-\cot(B \omega)|=O(e^{-2 min(A,B){\rm Im}~\omega})
\end{equation}
Thus on the upper and lower parts of the contour (${\rm Im}~z=\pm
L_{N}$) the integral is bounded by:
\begin{eqnarray}
\left|\int_{0}^{L_{N}}(iL_{N}+z)F_{l}(iL_{N}+z) dz\right|&\leq&
\int_{0}^{L_{N}}|iL_{N}+z|12 R e^{-4 a L_{N}} dz\nonumber\\&\leq&
18 L_{N}^{2}R e^{-4 a L_{N}}  \label{estimate}
\end{eqnarray}
The part of the contour with ${\rm Re}~z=L_N$ can be evaluated as
follows. First we prove that integrals of the form
\begin{equation}
\int_{-L}^{L} e^{-\sigma(L+ix)}(L+ix)\cot(kL+ikx)dx
\end{equation}
decay exponentially as $L\rightarrow\infty$ as long as we avoid
the poles (i.e. $kL\neq n\pi$). Choose
$L_{N}=\frac{N\pi}{k}+\frac{\pi}{2k}$; then we have
\begin{eqnarray}
&&\left|\int_{-L_{N}}^{L_{N}}
e^{-\sigma(L_{N}+ix)}(L_{N}+ix)\cot(kL_{N}+ikx)dx\right|=\nonumber\\
&& e^{-\sigma L_{N}}\left|\int_{-L_{N}}^{L_{N}} e^{-i\sigma
x}(L_{N}+ix)\tanh(kx)dx\right|\leq 3 L_{N}^{2} e^{-\sigma L_{N}}
\label{estimateside}
\end{eqnarray}
This is enough to ensure exponential decay for any $kL\neq n\pi$,
since we can shift the right side from $L_{N}$ to
$L_{N}-\frac{\pi}{2k}< L
< L_{N}+\frac{\pi}{2k}$ at a ``cost" of less then $\frac{\pi}{k}L_N^{2} e^{-\sigma
L_{N}}$. Combining estimates (\ref{estimate}) and
(\ref{estimateside}) we see that integration over the part of the
contour with ${\rm Re}~z>0$ decays exponentially like  $e^{-\sigma
L_{N}}$. In the limit $
 L_{N}\rightarrow\infty$ we obtain the desired equality:
\begin{equation}
\sum_{n=1}^{\infty} \omega_{(n,l)} e^{-\sigma
\omega_{(n,l)}}=\int_{-i\infty}^{i\infty}dz e^{-\sigma z}z {d\over
dz}\ln \Delta_{l}(z).
\end{equation}
In the limit of large $R$, we have on the imaginary axis (${\rm
Re}~\omega=0$)
\begin{equation}
\Delta_l^{(3)}\Delta_l^{(4)}(\omega)=\frac{1}{2\omega}
\sin(2\omega R-l \pi) {h_l}^{(1)}(\omega a){d\over d\omega}[\omega
{h_l}^{(1)}(\omega a)],
\end{equation}
where
\begin{equation}
h_l^{(1)}(\omega)=j_{l}(\omega)+in_{l}(\omega).
\end{equation}
After summation over the $l$ index and some algebraic manipulation
we obtain \cite{nesterenko97}:
\begin{eqnarray}
E_c=-{1\over
\pi}\lim_{\sigma\rightarrow0}\lim_{b\rightarrow\infty}\sum_{l=1}^{\infty}
(l+{1\over 2})\int^{\infty}_0 d\omega &e^{-i\sigma \omega} \omega
{d\over d\omega}\biggr(ln\big[1-{(s_{l}e_{l}(a
\omega))'}^2\big]\nonumber\\&-ln\big[1-{(s_{l}e_{l}(b
\omega))'}^2\big]\biggr)
\end{eqnarray}
in agreement with (3). Note that the usual derivation of (3)
involves a rescaling $x=a\omega$. In our derivation we avoided
this step and as a result, were able to identify the term in the
$\omega$-integral which is independent of the radius, and hence
makes (3) divergent.

\bigskip

\section{Expansion in $\xi^{2}$: first order}
In the previous section we reviewed the derivation of the
expression for the Casimir energy of a conducting sphere. The
Casimir energy of a dielectric ball, under the condition
$\sqrt{\epsilon\mu}=\sqrt{\epsilon'\mu'}=1$
\cite{nesterenko97,brevik98} can be calculated at no further cost
by repeating essentially the same steps as in the conducting case.
Furthermore, it was shown \cite{bordag99}, that calculations in
this case, will yield the same result independent of the
regularization used. From now on this will be the setting, with
$\xi=1$ (4) corresponding to the previous section. We now show
that the sum over the angular index $l$ in (3) can be carried out
before doing the integration over the frequencies. We assume
$\xi^2$ small, and expand the logarithm in (3) in powers of
$\xi^2$. The first term in this expansion is
\begin{equation}
-\xi^2 \sum_{l=1}^\infty (l+\frac{1}{2})x\der{x}{(s_{l}e_{l})'}^2
\label{sum1}.
\end{equation}
    It turns out that the sum over $l$ in (\ref{sum1}) can be done
exactly. To this end, note the following identity, which can
easily be obtained from the expansion of the Helmholtz propagator
in spherical coordinates \cite{ryzhik} using relations (5) and
(6),
\begin{equation}
D(x,\rho)\equiv\frac{\tilde{r} r xe^{-x
\rho}}{\rho}=\sum_{l=0}^\infty(2l+1)s_{l}(x \tilde{r})e_{l}(x r
)P_{l}(\cos\theta)
\end{equation}
\begin{equation}
\rho=\sqrt{\tilde{r}^2+r^2-2\tilde{r}r \cos\theta}
\end{equation}
where $\theta$ is the angel between two vectors of lengths $r$ and
$\tilde{r}$.

    The following can, of course, be carried out for arbitrary
$\tilde{r}$ and $r$ . This should be useful in the more general
case where $\sqrt{\epsilon\mu}\neq\sqrt{\epsilon'\mu'}$, since the
expressions appearing in that case involve combinations of
$s_{l}(x')e_{l}(x)$ where $x=\sqrt{\epsilon\mu}\omega$ and
$x'=\sqrt{\epsilon'\mu'}\omega$ \cite{brevik99}, but in the
present case $r=\tilde{r}=1$ is just what we need. Using the
orthogonality relations of the $P_{l}$'s
\begin{equation}
\int_{-1}^{1}P_{l}(x)P_{j}(x)dx=\delta_{lj}\frac{2}{2l+1}
\label{ortho}
\end{equation}
and substituting d$\cos\theta=-\rho d\rho$ we obtain the identity
\begin{equation}
\sum_{l=0}^\infty (2l+1){(s_{l}e_{l})'}^2 =\frac{1}{2}\int_{0}^{2}
(\frac{\partial}{\partial x}{D(x,\rho))}^2 \rho d\rho
\end{equation}
Thus the sum over $l$ in (\ref{sum1}) can be carried out, as
promised, and we find
\begin{eqnarray}
&-\frac{\xi^2}{2}(\frac{1}{2}\int_{0}^{2}
x\der{x}(\der{x}{D(x,\rho))}^2 \rho d\rho -
x\der{x}{(s_{0}e_{0})'}^2)=\nonumber\\&
\frac{\xi^2}{2}(\frac{1}{2}-\frac{1}{2}e^{-4x} (1-2x)^2+4xe^{-4x})
\label{sum2}
\end{eqnarray}
We note the appearance of a term $\frac{\xi^2}{4}$ which is
independent of $x$ and thus causes (3) to diverge. We dispose of
it by subtracting from (\ref{sum2}) the appropriate term
corresponding to a radius $b$ and taking the limit
$b\rightarrow\infty$, and obtain the density
\begin{equation}
G(\omega)\equiv\frac{\xi^2}{8\pi}e^{-4a|\omega|}(1+4a|\omega|+4a|\omega|^2)
\end{equation}
Thus, finally the first term in the $\xi^2$ expansion of (3) can
be analytically derived. The energy to this order is
\begin{equation}
E=\int_{-\infty}^{\infty}{d\omega G(\omega)}=\frac{\xi^2}{2
a}\frac{5}{16 \pi}+O(\xi^4)=\frac{\xi^2}{2 a}0.0994718+O(\xi^4).
\end{equation}
$\xi=1$ corresponds to the case of a conducting sphere. It turns
out that this result has already been derived for a conducting
sphere, by Balian and Duplantier \cite{balian78} using a multiple
scattering formalism, which may be applied in the case of
conducting boundaries of arbitrary shape. Using a Green's function
method, Milton, DeRaad and Schwinger \cite{milton78} evaluated the
Casimir Energy for a conducting sphere, Numerically, to be
$\frac{1}{2 a}0.092353$. This is in accordance with our
calculation, since the next term in the $\xi^2$ expansion of (3)
may be evaluated using the Debye asymptotic expansion of Bessel
functions to be $-\frac{1}{2 a}0.007$. However, for a dilute
sphere ($\xi<<1$) our result is the exact one. Note that our
result was obtained using only subtraction of energies, which is
in the definition of the Casimir energy. No further regularization
of the expression (\ref{sum1}) was necessary.
\bigskip
\section{An integral expression for higher orders in $\xi^{2}$}

    Generalization of the method we developed in the previous section
to higher orders in $\xi^2$, is possible. To this end, we derive a
general formula for sums of the form $\sum_{l}(l+1/2)a_{l}^{n}$,
for arbitrary power $n$, and coefficients $a_{l}$, when the
function $f(x)=\sum_{l}(l+1/2)a_{l}P_{l}(x)$ is known. Let us
first introduce the following transform
\begin{equation}
f(x)\rightarrow \widehat{f}(x,y)=\frac{1}{\pi} \int_{0}^{\pi}dt
f\big(xy-\sqrt{(1-x^2)(1-y^2)} \cos t\big)
\end{equation}
This transform has the simple property that if $f(x)=\sum_{l}
a_{l} P_{l}(x)$  then
\begin{equation}
\widehat{f}(x,y)=\sum_{l} a_{l} P_{l}(x) P_{l}(y)
\label{transform}
\end{equation}
To see this, we write $x=\cos\alpha$, $y=\cos\beta$ and consider
the following well known addition formula:
\begin{equation}
P_{l}(\cos\gamma)=P_{l}(\cos\alpha)P_{l}(\cos\beta) + 2
\sum_{m=1}^{l}P_{l}^{m}(\cos\alpha)P_{l}^{m}(\cos\beta)\cos(m(\theta-\varphi))
\end{equation}
where $\gamma$ is the angle between two unit vectors with
spherical coordinates $(\alpha,\theta)$ and $(\beta,\varphi)$,
that is
\begin{equation}
\cos\gamma=\cos\alpha\cos\beta+\sin\alpha\sin\beta\cos(\theta-\varphi)
\end{equation}
setting $\theta=t$ and $\varphi=0$, we obtain:
\begin{eqnarray}
 &f(\cos\alpha\cos\beta-\sin\alpha\sin\beta\cos t)=\sum_{l}  a_{l} P_{l}(\cos\gamma)
=\nonumber \nonumber\\& \sum_{l} a_{l} \big[P_{l}(\cos\alpha)
P_{l}(\cos\beta)+2
\sum_{m=1}^{l}P_{l}^{m}(\cos\alpha)P_{l}^{m}(\cos\beta)\cos(mt)\big]
\end{eqnarray}
Integrating over $t$, we eliminate all the terms containing $\cos
t$, and are left with $\sum_{l} a_{l}P_{l}(x)P_{l}(y)$ as claimed.
Using (\ref{transform}) and ({\ref{ortho}) repeatedly, we obtain
\begin{equation}
\sum_{l=0}^{\infty}(l+1/2)a_{l}^{n}=\int_{-1}^{1}\prod_{j=1}^{n-1}dx_{j}\big(
f(x_{1})\underbrace{\widehat{f}(x_{1},x_{2})\cdot...
\widehat{f}(x_{n-2},x_{n-1})}_{(n-2) terms} f(x_{n-1})\big)
\label{sum3}
\end{equation}
Thus, for example, we can cast the $\xi^4$ term in the expansion
of (3) in the form:
\begin{equation}
\frac{-\xi^4}{32\pi^3a}\Big(\int_{0}^{\infty}dx\int_{-1}^{1}
d\cos\theta_{1}d\cos\theta_{2}d\cos\theta_{3}\int_{0}^{\pi}dt_{1}dt_{2}
\prod_{i=1}^{4}\der{x}D(x,\rho_{i})-\pi^2\Big) \label{xi4}
\end{equation}
where
\begin{eqnarray}
\rho_{1}=\sqrt{2-2\cos\theta_{1}}\,\,\, ;\,
\rho_{2}=\sqrt{2-2(\cos\theta_{1}\cos\theta_{2}-\sin\theta_{1}
\sin\theta_{2}\cos t_{1})} \nonumber \\
\rho_{3}=\sqrt{2-2(\cos\theta_{2}\cos\theta_{3}-\sin\theta_{2}
\sin\theta_{3}\cos t_{2})}\,\,\, ; \,
\rho_{4}=\sqrt{2-2\cos\theta_{3}} \label{xi}
\end{eqnarray}
It is not difficult to check that (\ref{xi4}) indeed converges
but, unfortunately, we have not succeeded in integrating it
analytically.

\section{Discussion}

    In this paper we developed and used a novel method appropriate for
calculations of the Casimir energy for spherical boundary
conditions. Although the use of the asymptotic Debye approximation
is very convenient, we have shown that it is possible to make a
direct summation over angular modes using the expansion of the
Helmholtz propagator in spherical harmonics. This method can be
applied to obtain closed integral representations for any order in
the $\xi^2$ expansion of the Casimir energy (3). The $\xi^2$ order
was explicitly calculated to be $\frac{5\xi^2}{32a\pi}$, this
result is within $0.13\%$ of results obtained via the Debye
approximations \cite{brevik98}, demonstrating their high accuracy.
Higher orders in $\xi^2$ however, still remain in integral form.

    Our method should be applicable to other sums of the type
(\ref{sum1}) and variations on them \cite{brevik99}, which are
common in the calculations for spherical boundaries, in order to
get more regularization independent results. The fact that
summation using angular integrals over the propagator of the
Helmholtz equation through (\ref{sum3}) can be done to all orders,
reflects the relation between the mode summation technique and
calculations using Green's function. This can be seen for the case
of conducting boundaries in the multiple scattering formalism
\cite{balian78}. It would be illuminating to reveal the
connections between our method and other methods of calculation in
the dielectric case, involving summation over dipoles, such as
calculations of the Casimir energy using the statistical mechanics
 partition function as was performed by H\o ye and
Brevik \cite{hoye99}.

\section*{acknowledgments}

    I am grateful to M. Revzen for interesting me in this problem and
for his insight and to J. Feinberg and A. Mann for their advice
and comments. I also wish to thank A. Elgart for useful
discussions and Professor I. Brevik for careful reading of the
manuscript and helpful comments. 
\end{document}